\documentstyle[preprint,aps]{revtex}

\begin{document}
\preprint{\font\fortssbx=cmssbx10 scaled \magstep2
\hbox to \hsize{
%\special{psfile=/NextLibrary/TeX/tex/inputs/uwlogo.ps
%			      hscale=8000 vscale=8000
%			       hoffset=-12 voffset=-2}
%\hskip.5in \raise.1in
\hbox{\fortssbx University of Wisconsin - Madison}
\hfill$\vcenter{\hbox{\bf MADPH-96-948}
          \hbox{July 1996}}$ }
}

\draft
\title{\vspace*{.25in} Gamma ray astronomy with muons}
\author{Francis Halzen$^1$, Todor Stanev$^2$ and Gaurang B. Yodh$^3$}
\address{
$^1$ Department of Physics, University of Wisconsin, Madison, WI 53706\\
$^2$ Bartol Research Institute, University of Delaware, Newark, DE 19716\\
$^3$ Department of Physics and Astronomy, University of California,
Irvine, CA 92715}
\maketitle
\begin{abstract}\baselineskip15pt
 Although gamma ray showers are muon-poor, they still produce a number
 of muons sufficient to make the sources observed by GeV and TeV
 telescopes observable also in muons. For sources with hard gamma
 ray spectra there is a relative `enhancement' of muons from gamma ray
 primaries as compared to that from nucleon primaries.
 All shower gamma rays above the photoproduction threshold 
 contribute to the number of muons $N_\mu$, which is thus 
 proportional to the primary gamma ray energy.  With  gamma ray
 energy 50 times higher than the muon energy and a probability
 of muon production by the gammas of about 1\%,  muon detectors
 can match the detection efficiency of a GeV satellite detector
 if their effective area is larger by $10^4$.
 The muons must have enough energy for sufficiently accurate
 reconstruction of their direction for doing astronomy.
 These conditions are satisfied by relatively shallow neutrino
 detectors such as AMANDA and Lake Baikal and by gamma ray detectors
 like MILAGRO. TeV muons from gamma ray primaries, on the other hand,
 are rare because they are only produced by higher energy
 gamma rays whose flux is suppressed by the decreasing flux at the
 source and by absorption on interstellar light.  We show that there
 is a window of opportunity for muon astronomy with the AMANDA,
 Lake Baikal and MILAGRO detectors.
\end{abstract}
\thispagestyle{empty}

\section{Introduction}

  Instruments exploiting the air Cherenkov technique have extended the
 exciting astronomy revealed by the Compton G(amma) R(ay) O(bservatory)
 into the TeV energy range\cite{Weekes}. The photon spectra of the Crab
 nebula, the pulsar PSR 1706-44 and of the active galaxy Markarian 421,
 observed by satellite experiments at GeV energies, are known to extend
 to higher energy, e.g.\ well above 10~TeV for the Crab. More
 interestingly, the galaxy Markarian 501, recently detected in the TeV
 range, is not a confirmed GeV gamma ray source.

  High energy gamma rays produce muons in the Earth's atmosphere
 that can be detected and reconstructed in relatively shallow
 underground neutrino detectors such as the AMANDA and Lake Baikal,
 which are positioned at a modest depth of order 1~kilometer~\cite{GHS}
 or in a surface detector like MILAGRO. The neutrino detectors are
 sensitive to muon energies of a few hundreds of GeV and the Milagro
 telescope detects muons above 1.5 GeV, well below the TeV thresholds
 of deep underground detectors. Although muons from such sources
 compete with a large background of down-going cosmic ray muons,
 they can be identified provided the detectors have a sufficient
 effective area. Unlike air Cherenkov telescopes, muon detectors
 cover a large fraction of the sky with a large duty cycle,
 e.g.\ 100\% efficiency for more than a quarter of the sky in the case of
 the AMANDA detector with a South Pole location. The advantage is
 considerable  in studying the emission from highly variable sources.
 Useful results may, possibly, be obtained with the partially deployed
 instruments, even before they achieve the necessary up-down
 discrimination to identify neutrinos. Moreover, background multi-muon
 bundles, which are difficult to reconstruct, can be conveniently
 rejected without suppression of the predominantly single-muon
 gamma ray signal.

  In this paper we demonstrate how these muon energy ranges provide
 a window of opportunity for muon astronomy. The muons are sufficiently
 energetic to leave tracks that can be adequately measured by the Cherenkov
 technique. The direction of the parent photon can be inferred with
 degree accuracy in the case of AMANDA and Lake Baikal and a few degree
 accuracy in the case of MILAGRO.

  Detected muons originate in gamma showers with some 50 times higher
 energies.  Such gamma rays have been detected, at least from two
 galactic and two extra-galactic sources, by air-Cherenkov telescopes.
 A multi-TeV air shower will produce a 100~GeV muon with a
 probability of order 1\%\cite{HHS}, sufficient to observe the
 brightest sources using relatively modest-size detectors with
 effective area of order 1000~m$^2$ or more. The probability
 that such photons produce TeV-energy muons which trigger the deep
 underground detectors, such as those in the Gran Sasso tunnel,
 is small. TeV muons are produced by photons  of several tens of
 TeV energy and above. Their rates are suppressed and, more
 importantly, it is not clear whether the source spectra
 extend far beyond the TeV region. They most likely do not.
 In the case of galactic sources, such as the Crab supernova remnant,
 the current thinking is that the high energy photons are produced
 by inverse Compton scattering of electrons accelerated by the
 pulsar\cite{HdeJ}. Such a purely electromagnetic accelerator is
 unlikely to produce photons far beyond the observed spectrum which
 extends to 10~TeV. While the vast majority of GeV gamma ray sources
 display a $E^{- (\gamma+1)}$ energy spectrum with $\gamma
 \simeq 1$\cite{2EC}, the steepening slope of the Crab spectrum may
 provide us with a glimpse of a steep cutoff not far beyond the reach
 of the data; see Table~1. Another type of GeV~\cite{Esp}, and  possibly
 TeV, galactic gamma ray sources are the supernova remnants, presumably 
 the acceleration sites of the galactic cosmic rays. The recent
 observation of hard X--rays at SN1006~\cite{SN1006} is consistent
 with the acceleration of electrons to 100 TeV. Extragalactic
 sources such as the active galaxy Markarian 421 may, on the other
 hand, be true high energy accelerators producing protons with
 sufficient energy to account for the high energy cosmic ray spectrum
 which extends beyond $10^{20}$~eV. Even in this case, the existence
 of higher energy photons is not guaranteed because they may be
 absorbed in the source, or on the interstellar infrared and microwave
 background\cite{SS,RJPTS}. Nearby active galactic nuclei in
 the local cluster or the super-galactic plane, with redshift
 less than 0.03 or so, represent promising sources in this respect.
 Examples are listed in Table~1 where, as usual, we have parameterized
 the gamma ray flux as
\begin{equation}
{dN_\gamma\over dE_\gamma} = {F_\gamma   \over E^{(\gamma+1)}} 10^{-12} \rm\,
cm^{-2} \, s^{-1}\;.
\end{equation}

  Throughout this paper energies are in TeV units. The high energy
 luminosity of the source is described by the parameter $F_{\gamma}$
 which in the EGRET catalog\cite{2EC} denotes the flux of photons
 above 100~MeV in units of $10^{-8}$ photons per cm$^2$ per second.
 For flat ($\gamma = 1$) spectra the same number will apply to the
 flux of TeV gamma rays in units $10^{-12}$. Most EGRET sources with
 measured energy spectrum have $\gamma \leq 1$. For the Crab supernova
 however the TeV flux is reduced by one order of magnitude with
 $F_{\gamma} = 20$ in the TeV region\cite{WhippleC}. This number is
 bracketed by the 7(50) Markarian 421 flux in its low (high)
 state\cite{WhippleM}. So, interestingly, galactic and nearby
 extra-galactic sources produce comparable photon fluxes at Earth
 despite the $10^5$ ratio of their distances. Sources such as the
 Vela pulsar, the galactic center and the cluster of four unidentified
 gamma ray sources in the direction of the spiral arm near Cygnus,
 may be TeV gamma ray emitters brighter by more than
 one order of magnitude. We refer to Table~1 for a partial list.

\section{Muon Rates from Gamma Ray Sources}

  Gamma rays initiate atmospheric cascades of mostly electrons and
 photons, but also some muons. Muons originate from the decay of
 pions which are photoproduced by the shower photons. The number of
 muons with energy above $E_{\mu}$ in a shower initiated by a photon
 of energy $E_{\gamma}$ was computed some time ago. For $E_{\mu}$
 in the range from 0.1 to 1~TeV the number of muons in a photon
 shower can be parameterized as
\begin{equation}
N_\mu (E_\gamma, > E_\mu) \cong {2.14\times 10^{-5} \over \cos\theta}
{1\over (E_\mu/\cos\theta)} \left[ E_\gamma \over (E_\mu/\cos\theta) \right]\,.
\end{equation}
  Energy units are TeV and the parameterization is adequate for
 $E_{\gamma}/E_{\mu} \geq 10$. The estimate is conservative and below
 the rate of muons obtained by Bhattacharyya\cite{Bhat}, who updated the
 calculations of reference\cite{HHS} taking into account the latest
 measurements of the high energy photoproduction cross section at HERA.
 TeV-muons are also produced by muon pair production and the decay of
 charm particles. For the lower energy muons considered here these additional
 sources can be neglected\cite{HHS}. For muons of energy lower than
 0.1 TeV, more complete formulas of \cite{HHS} must be used and
 probability of muon decay incorporated. The muon flux produced by a gamma
 ray source is obtained by convolution of Eqs.~1 and~2:
\begin{eqnarray}
N_\mu ({>}E_\mu) &=& \int_{E_{\gamma\,{\rm min}}}^{E_{\gamma\,{\rm max}}}
dE_\gamma
{F_\gamma 10^{-12} \over E_\gamma^{\gamma+1}}\,{2.14\times 10^{-5} \over E_\mu}
\left( E_\gamma \cos\theta \over E_\mu \right)\;, \\
&\simeq& 2\times 10^{-17} {F_\gamma\over \cos\theta}
{1\over (E_\mu/\cos\theta)^{\gamma+1}}\, \ln\!\!\left(E_{\gamma\,{\rm max}}
\over E_{\gamma\,{\rm min}} \right) f \;.
\end{eqnarray}
  Here $E_{\mu}$ is the vertical threshold energy of the detector,
 e.g.\ 0.18~TeV for the AMANDA detector. $\theta$ is the zenith
 angle at which the source is observed. It is, conveniently,
 time-independent for the AMANDA detector with a South Pole location.
 Photons with energy ranging  from a minimum energy
 $E_{\gamma\,{\rm min}} \simeq 10 \times E_{\mu} / \cos\theta$
 to the maximum energy of the source $E_{\gamma\,{\rm max}}$
 contribute to the production of the observed muons. The highest
 energy photons dominate. For this reason the muon flux depends
 critically  on the high energy flux of the source. The factor $f$
 is a correction factor which can be parametrized as
\begin{equation}
f = \left(E_\mu/\cos\theta\over 0.04\right)^{0.53} \;.
\end{equation}
 The flux of muons varies with the vertical threshold of the
 detector as $E_{\mu}^{-(\gamma+1)}$. This behavior is only
 approximate and assumes that the integrand in Eq.~3 spans many
 decades of the $E_{\gamma}^{-2}$ spectrum between
 $E_{\gamma\,\rm min}$ and $E_{\gamma\,\rm max}$. Otherwise,
 the dependence is moderated, an effect which is described by
 the factor $f$. In the end our parameterization reproduces the
 explicit Monte Carlo results\cite{HHS}.

  A corresponding calculation has to be carried out to
 determine the background of muons from primary cosmic rays.
 For the AMANDA detector the gamma ray signal has to be extracted
 from a background of cosmic ray muons which is empirically
 $3.3 \times 10^{-6}$ muons per cm$^2$ per second per steradian
 and falls with zenith angle as $\cos\theta^{2.8}$ at a detector
 depth of 1~kilometer water-equivalent\cite{Baikal}. For the MILAGRO
 detector the cosmic ray muon flux has a cos$(\theta)^2$ angular
 distribution and the rate is $\sim$ 0.01 muons in the same
 units\cite{yodh}.

  For AMANDA the relevant number is the background
 muons in a pixel of $\delta \times \delta$ degrees which is given by
\begin{equation}
N_\mu^{\rm back} (\rm m^{-2} \, yr^{-1}) \simeq 325\cos\theta^{2.8} \delta^2
\;.
\end{equation}
  As previously mentioned the background includes some fraction of
 multi-muon events. Rejecting multi-muon events not only improves
 signal-to-noise, it should improve angular resolution which is often
 degraded by the poor reconstruction of complex muon bundles initiated
 by high energy cosmic ray muons.  For MILAGRO, which detects low energy
 muons the mean angular accuracy achievable is dominated by the spreading
 due to transverse momentum kick given to the pions at production and is
 of the order of 3$^{\circ}$. The effective area of MILAGRO detector for
 muons is 1.5$\times 10^{3}$~m$^2$.

\section{Summary of Results and Examples}

\subsection{The AMANDA Telescope:}
  Our results can be conveniently summarized as follows. The
 number of events per year in a detector of effective area
 $A$ m$^2$ and vertical threshold of 0.18~TeV is given by
\begin{equation}
N_\mu ({\rm yr^{-1}}) = 6.7\times 10^{-6} {F_\gamma\over\cos\theta}
{1\over(E_\mu/\cos\theta)^{\gamma+1}} \,
\ln\!\!\left(\cos\theta E_{\gamma\,{\rm max}} \over 10E_\mu\right) f A \;.
\end{equation}
 We recall that all energies are in TeV units.
 The signal-to-noise, defined as the number of events divided
 by the square root of the number of background events in a
 pixel of $\delta \times \delta$ degrees, depends on detector
 area and zenith angle as
\begin{equation}
S\Big/\sqrt N \sim {\sqrt A\over \cos\theta^{0.9} \delta} \;.
\end{equation}
 The formula simply expresses that signal-to-noise is
 improved for increased area, better resolution and for sources
 observed at large zenith angle where the cosmic ray background
 muon rate is reduced.

 To demonstrate the power of a relatively shallow neutrino detector
 as a gamma-ray telescope we start with an optimistic, though not
 unrealistic example. We take the Vela pulsar with $F_{\gamma} = 932$
 and $\theta = 45^\circ$ at the South Pole. This assumes that the
 source flux is not cut off at high energy and the spectral index
 $\gamma$ is 1. We will however assume that $E_{\gamma\,\rm max}$
 is only 10~TeV (the muon flux is increased by a factor 4.4 if the
 spectrum extends to 1000~TeV). For a nominal detector with effective
 area $10^4$~m$^2$ area and $\delta = 1^\circ$ angular resolution
 we obtain 5000~events per year above $E_\mu \;=\;
 {{ {\rm 180 GeV}} \over {\cos({\rm 45}^\circ)}}=255$~GeV
 on a background of $1.2 \times 10^6$ or an $S/\sqrt N$ ratio
 in excess of~4.

  For the blazar Markarian~421 the TeV-flux, averaged between the
 high and low state, corresponds to $F_{\gamma} \simeq 35$.
 Our nominal detector should collect 442 events from such a source
 at a zenith angle of 60$^\circ$ per year assuming
 $E_{\gamma\,\rm max} =$(100~TeV). This would give a $S/\sqrt N$
 ratio of 0.6.

  As a final example we propose the 4 sources in a 5$^\circ$
 by 3$^\circ$ declination/right ascension bin in the direction
 of the spiral arm in Cygnus. For this cluster of sources
 $F_{\gamma} = 335$ and $\theta = 60^\circ$; we will assume
 that $E_{\gamma\,\rm max} = 100$~TeV. A $10^4\rm~m^2$ detector
 will collect 3700~events on a background of 18~million in the
 bin containing the sources for $S/\sqrt N $ close to 1.
 No precise reconstruction is required. A similar argument
 should apply to the galactic plane.

\subsection{ The MILAGRO Telescope}
   As the muon threshold of MILAGRO is only 1.5 GeV, it can
 probe lower energy gamma rays than that for the neutrino
 telescopes. It has sensitivity to an energy range which
 overlaps that of Whipple telescope. The signal that can be
 observed depends sensitively on the cutoff of the gamma ray
 spectrum in the 100~GeV to 10~TeV energy range. As mentioned
 earlier, the 3$^{\circ}$ intrinsic angular resolution requires
 that one must use a 4.7$^{\circ}$ bin size to collect $\sim$~70\%
 of the gamma ray events. The background cosmic ray flux is
 then $\sim$ 930~Hz. The sensitivity of MILAGRO muons for gamma ray
 searches is shown in Table~2 for several sources, which gives
 the time in seconds required to obtain a 5.5$\sigma$ significance
 result, for different cutoff energies of the primary gamma
 spectrum. The first column identifies the source, the second
 and third define the source parameters. The fourth column
 gives the maximum energy cutoff that is assumed for the
 incident gamma ray spectrum, fifth the expected rates for
 signal, and the last column gives the time in seconds required
 to obtain a 5.5$\sigma$ observation.

  Table 2 shows that Geminga may be observed even if the cutoff
 is only 300 GeV, in about a year. If arrival times of events
 can be put into the data stream for events coming from
 Geminga, period analysis could be applied to improve the
 signal to noise ratio and probably lower the cutoff energy
 down to 50~GeV.

  Halzen 4 is a set of 4 Egret sources, described in the previous
 subsection, which are within about 2 degrees of each other
 with a reasonably flat spectra. It is observable, if the spectrum
 extends to energies above a TeV.

  Finally, we note that stronger GRBs should be  observable,
 even with a 1.5 energy spectrum, if they last as long as the
 times indicated in the last column. For the GRBs with flatter
 spectra the possibility of observation is enhanced. The
 dependence on cutoff energy is also indicated in the table.
 For several cases one can a see a signal even if the cutoff is
 as low as 100 GeV. Thus, at least for GRBs the muon observation
 allows us to extend the energy range of Milagro to lower cutoff
 energies.

{\bf\noindent Acknowledgments.}
 We thank Christian Spiering for a careful reading of the
 manuscript. This research was supported in part by the U.S.~Department of
 Energy under Grants No.~DE-FG02-95ER40896 (FH) and
 DE-FG-91ER40626 (TS) and in part by the
 University of Wisconsin Research Committee with funds granted by
 the Wisconsin Alumni Research Foundation. The research of GBY is
 supported in part by the Physics Division of NSF.

\bigskip
\begin{table}%%[h]
\centering
\tabcolsep.75em
\caption{ A partial list of potential VHE $\gamma$--ray sources based on the
2nd EGRET catalog. Groups of sources that are difficult to resolve are
combined. The position for such sources are averaged with the EGRET
$\gamma$--ray flux and a solid angle (ster) for the group is given.
$F_\gamma$ is the average number of photons above 100 MeV in units
of $10^{-8}\rm\,cm^{-2}\,s^{-1}$, $\gamma$ is the spectral index. The check
marks indicate sources observed by TeV Cherenkov telescopes.}
\smallskip
\begin{tabular}{rrccccc}
\hline
RA& Dec& $F_\gamma$ & $\gamma$& $\Delta\Omega$&
Source& VHE? \\
\hline
128.8& $-$45.2& 932$\pm$8& 0.6--0.9& ---& Vela pulsar& $\surd$\\
98.5& 17.8& 374$\pm$ 5& 0.4--0.7& ---& Geminga pulsar& $\surd$\\[2mm]
306.1& 39.2& 335$\pm$ 9& ---& $3.3\times10^{-3}$&
$\left\{\parbox[c]{.875in}{J2019+3719\\ J2020+4026\\ J2026+3610\\
J2033+4122}\right.$& ---\\[14mm]
266.8& $-$29.8& 218$\pm$ 10& ---& $2.7\times10^{-4}$&
$\left\{\parbox[c]{.875in}{J1746$-$2852\\ J1747$-$3039}\right.$& ---\\[2mm]
184.56& $-$5.8& 212$\pm$3& 1.2--1.5& ---& Crab pulsar& $\surd$\\
257.4& $-$44.9& 144$\pm$ 5& ---& ---& PSR1706$-$44& $\surd$\\[2mm]
217.7& $-$23.4& 121$\pm$ 9& ---& $3.6\times10^{-4}$&
$\left\{\parbox[c]{.875in}{J1801$-$2312\\ J1811$-$2339}\right.$& ---\\[2mm]
213.2& $-$62.2& 103$\pm$ 9& ---& ---& J1412$-$6211& ---\\
155.4& $-$58.6& 99$\pm$ 5& ---& ---& J1021$-$5835& ---\\
\hline
\end{tabular}
\end{table}
\bigskip
\begin{table}%%[h]
\centering
\tabcolsep.75em
\caption{Sensitivity of MILAGRO muons to sources}
\smallskip
\begin{tabular}{l|ccccc} \hline
Source & $F_{\gamma}$&Diff slope & $E_{\rm max}$ & $S$ (Hz) &  Time (sec) \\
       &             & $\gamma$  & GeV  &	& to 5.5$\sigma$ \\ \hline
Geminga & 374        & 1.5       & 10$^{2}$& .013 &  1.7x10$^{8}$   \\
Geminga & 374        & 1.5       &3x10$^{2}$&.04 &  1.4x10$^{7}$   \\
Geminga & 374        & 1.5       & 10$^{3}$& .15 &  1.2x10$^{6}$\\ \hline
Halzen 4 & 336       & 1.7       & 10$^{4}$& .11 &  2.1x10$^{6}$ \\
Halzen 4 & 336       & 1.7       &5x10$^{3}$& .08 &  4.4x10$^{6}$ \\
Halzen 4 & 336       & 1.7       &3x10$^{3}$& .06 &  7.5x10$^{6}$ \\
Halzen 4 & 336       & 1.7    &1x10$^{3}$& .03 &  2.8x10$^{7}$ \\ \hline
   AGN  & 100       & 1.7    & 10$^{4}$& .035 & 2.4x10$^{7}$\\\hline
GRB     & 6.6e6     & 2.5       &1x10$^{3}$&  16.7 &  102           \\
GRB     & 6.6e6     & 2.5       &5x10$^{2}$& 14.8 &  131           \\
GRB     & 6.6e6     & 2.5       &3x10$^{2}$& 13.0 &  167           \\
GRB     & 6.6e6     & 2.5       &2x10$^{2}$&  8.4 &  404      \\ \hline
GRB     & 6.6e6     & 2.3       &1x10$^{3}$& 36.6 &   21          \\
GRB     & 6.6e6     & 2.3       &5x10$^{2}$& 31.3 &   29          \\
GRB     & 6.6e6     & 2.3       &3x10$^{2}$& 27.05 &  39          \\
GRB     & 6.6e6     & 2.3       &1x10$^{2}$& 16.4  &  106      \\ \hline
GRB     & 6.6e5     & 2.5       &1x10$^{3}$& 1.67 &   10210          \\
GRB     & 6.6e5     & 2.5       &5x10$^{2}$& 1.47 &   13114          \\
GRB     & 6.6e5     & 2.5       &3x10$^{2}$& 1.30 &    16788          \\
GRB     & 6.6e5     & 2.5       &1x10$^{2}$& .84 &   40455      \\ \hline
GRB     & 6.6e5     & 2.3       & 10$^{3}$ & 3.66 &    2125         \\
GRB     & 6.6e5     & 2.3       &5x10$^{2}$& 3.14 &   2899         \\
GRB     & 6.6e5     & 2.3       &3x10$^{2}$& 1.63 &   10597     \\  \hline
GRB     & 6.6e4     & 2.5       &1x10$^{3}$& .16 &  10$^{6}$       \\
GRB     & 6.6e4     & 2.0    &1x10$^{3}$& 1.32 &  1.6x10$^{4}$ \\ \hline
\end{tabular}
\end{table}

\end{document}